\documentclass[conference]{IEEEtran}
\IEEEoverridecommandlockouts
\usepackage{cite}
\usepackage{amsmath,amssymb,amsfonts}
\usepackage{algpseudocode}
\usepackage{graphicx}
\usepackage{textcomp}
\usepackage{makecell}
\usepackage[table]{xcolor}
\usepackage{hyperref}
\usepackage[inline]{enumitem}
\usepackage{csvsimple}
\usepackage{booktabs}
\usepackage{datatool}
\usepackage{array}
\usepackage{multirow}
\usepackage{lipsum}
\usepackage{afterpage}
\usepackage{tikz}
\usetikzlibrary{automata,arrows,positioning,calc}
\usepackage{caption}

\font\titlefont=ptmb7t at 35pt
\font\subtitlefont=ptmr7t at 21pt
\font\tinyfont=ptmr7t at 14pt

\def\BibTeX{{\rm B\kern-.05em{\sc i\kern-.025em b}\kern-.08em
    T\kern-.1667em\lower.7ex\hbox{E}\kern-.125emX}}

\usepackage{varwidth}
\DeclareCaptionFormat{myformat}{%
  \begin{varwidth}{\linewidth}%
    \centering
    #1#2#3%
  \end{varwidth}%
  }

\title{\titlefont{ The Universal Trust Machine }\\ \subtitlefont{ A survey on the Web3 path towards enabling long term digital cooperation through decentralised trust}\\ \tinyfont{ --- student project ---}}
\author{
\IEEEauthorblockN{Rohan Madhwal}
\IEEEauthorblockA{
Delft University of Technology\\
Delft, The Netherlands \\
R.Madhwal@student.tudelft.nl}
\and
\IEEEauthorblockN{Johan Pouwelse}
\IEEEauthorblockA{
Delft University of Technology\\
Delft, The Netherlands \\
J.A.Pouwelse@tudelft.nl}
}

\begin{document}

\maketitle
\thispagestyle{plain}
\pagestyle{plain}

\begin{abstract}
Since the dawn of human civilization, trust has been the core challenge of social organization. Trust functions to reduce the effort spent in constantly monitoring others' actions in order to verify their assertions, thus facilitating cooperation by allowing groups to function with reduced complexity. To date, in modern societies, large scale trust is almost exclusively provided by large centralized institutions. Specifically in the case of the Internet, Big Tech companies maintain the largest Internet platforms where users can interact, transact and share information. Thus, they control who can interact and conduct transactions through their monopoly of online trust. However, as recent events have shown, allowing for-profit corporations to harness so much power and act as gatekeepers to the online world comes with a litany of problems. While so far ecosystems of trust on the Internet could only be feasibly created by large institutions, Web3 proponents have a vision of the Internet where trust is generated without centralised actors. They attempt to do so by creating an ecosystem of trust constructed using decentralised technology. This survey explores this elusive goal of Web3 to create a ``Universal Trust Machine", which in a true decentralised paradigm would be owned by both nobody and everybody. In order to do so, we first motivate the decades-old problem of generating trust without an intermediary by discussing Robert Axelrod's seminal research on the evolution of cooperation in the iterated prisoner's dilemma. Next, we present the infrastructural and social challenges that a hypothetical Universal Trust Machine would have to overcome in order to enable long term cooperation in a decentralised setting. We proceed to present various reputation systems, all of which present promising techniques for encouraging trustworthy behaviour in a decentralised network through indirect reciprocity. After this, we discuss the family of emerging Distributed Ledger technologies whose secure transaction facilitating and privacy preserving techniques promise to be a good complement to the current limitations of vanilla reputation systems. Finally, we conclude by discussing a future roadmap for creating the desired Universal Trust Machine.
\end{abstract}


\section{Introduction} \label{intro}
Humans in a society rely on trust in every stage of their life, in every action they perform. Children trust that their parents will nurture and guide them, adults trust that their family and loved ones won’t deceive them. When crossing the street on a zebra crossing, we trust that motorists will obey the traffic laws, when buying items at the market, we trust in the quality of the goods being provided to us. Regardless of whether one believes that society is a function of divine order or of a social contract, trust between its members is the very fabric of its organising foundation. \cite{compModels} 

More generally, consider an agent, such as a human or a robot, who is required to use limited agency to navigate and take actions in a world with limited direct information available to it at any given moment. In such a world, trust is an important social heuristic that allows the agent to make wagers on the predictive benevolence of other agents. \cite{trustInSocialInt} 

Ecologist Garett Hardin defines trust as “encapsulated interest”, since it facilitates peaceful and stable social relations that form the basis of collective behavior and productive cooperation. Thomas Hobbes, considered by many to be one of the founders of modern political philosophy, argues that the natural state of humans is nasty and brutish, however, trust helps to convert that into something peaceful and efficient. In his book “A treatise of human nature”, enlightenment philosopher David Hume discusses the importance of trust to the functioning of a society. According to sociologist and philosopher Niklas Luhmann, trust effectively reduces complexity and risks, allowing for coordination with increased performance. \cite{luhmann} This is easy to understand intuitively since trusting individuals and groups reduces the effort one would spend in constantly monitoring the actions of others in order to verify their assertions. It is easy to conclude that a society without a notion of trust would find it hard to function effectively, or to exist at all. \cite{trustSocialCapital}

The growth of human civilization from small-scale hunter-gatherer societies to thriving economies of nation states is testament to the benefits provided by the growth of trust and cooperation inside societies. However, history reminds us that the requirement of trust for facilitating cooperation also leads to the growth of large centralized institutions since these institutions historically provided the best defense in economic transactions against the untrustworthy. \cite{cook2009whom}

While trust might be fundamental to cooperation in a society, underlying every social transaction is the desire to further one’s personal gain by abusing the trust of an unsuspecting opponent and defecting against the expected trustworthy action. \cite{compModels} For example, in a transaction where a merchant pre-pays a farmer for their produce at the end of the year, the farmer may be tempted to keep the payment and not provide the promised crops, or provide crops of a lower quality than was agreed upon. 

According to Margaret Levi, “good defenses make good neighbors”. Hence, the need for such defenses in economic transactions necessitated institutional bases of reaching agreement and resolving disputes that might result from them. Institutions that were able to provide third party enforcement in a transaction were hence able to ensure personal security and the security of the transaction. Thus, they were able to encourage cooperation and grow immensely as a result of their importance in doing so. \cite{cook2009whom}

However, allowing profit driven institutions to amass so much power comes with its own set of problems. The financial crisis of 2008 which was primarily attributed to failure of trusted institutions such as banks and other financial institutions has led to a growing distrust in such institutions. \cite{trustFinancialCrisis} This was most notably witnessed by the recent growth of blockchain technology and adoption of cryptocurrencies such as Bitcoin and Ethereum as a decentralised alternatives to large financial institutions.

The Internet is the most remarkable addition to how social capital can be built in the world. Collaborative work performed on the Internet is continuously changing how humans think about social interaction. To understand how trust is built on the Internet, it is worth considering the similarities and differences between trust on the Internet and trust in general. 

Since users on the Internet often possess virtually no knowledge about each other, all they can rely on is the immediate record of the other party's behaviour in past interactions with them to decide whether they can be trusted. However, this inability to directly judge different providers of services on the Internet is not very different from the general inability to directly judge the quality of services that are required in the real world, such as doctors or lawyers. Similar to the real world, providers of service on the Internet need to care not only about their current interaction, but also the result of the interaction on their future reputation. Hence, building and maintaining one's reputation by acting in a trustworthy manner is a requirement both in the real world and on the Internet. \cite{hardin2006trust}

On the other hand, the most notable difference between the two is caused by the Internet's unique capacity to allow collaboration and interactions at a global level. Take for example the case of buying an item from a local store, while doing so, trust is generally not an issue and most often, all that matters is the perceptible quality and price of the goods being provided in the store. However, buying an item from a seller on an online marketplace like eBay requires a markedly different level of trust to allow the transaction to occur. Since, in addition to simple quality and price of the advertised goods, the buyer would also require that reliable behaviour from the seller is guaranteed. Given two online sellers that sell the exact same item at the same price, a buyer would prefer the seller that has a large number of reviews/testimonials. Therefore, even in simple transactions, due to the global scale of the Internet, the risk of fraud is substantial and hence, additional methods of generating trust are required. \cite{hardin2006trust}

Even though the Internet was built on distributed protocols, large scale cooperation was consolidated around a few centralised services where social trust was created and enforced by large profit driven institutions. \cite{Korpal2022} Specifically, in two key functions of the web, web-publishing and discovery of content, technological institutions such as Google, Meta and Twitter slowly became curators and gatekeepers for the information being published on the Internet and people who were allowed to interact with it. As a result of this, the platforms accrued the power to control and own a large share of the information published and consumed on the Internet. 

Recently however, abuses of information and communication technology by such institutions for surveillance, spreading of disinformation and coercion of the public have come to light. Notable examples include Google's deepening involvement with Egypt's repressive government and Twitter enabling the Chinese government to promote disinformation on the repression of Uighurs. \cite{blockchainConfidenceMachine}

Such propensity of Big Tech organisations to abuse their ecosystems of trust for their own profit through privacy violations and misinformation is leading to a shift in the general attitude towards large centralised information platforms. The presence of large centralised authorities or platform owners to maintain and enforce trust in sociotechnical systems is increasingly being viewed more as a hindrance rather than a help. \cite{blockchainConfidenceMachine}

A growing alternative to the existing model of the platform driven Internet is the idea of Web3 which is motivated by the idea of using decentralised technologies such as blockchain. It is hard to exactly define Web3 since there is a lack of consensus even among researchers on what the idea of Web3 means. In section \ref{web3} we attempt to clearly define what Web3 refers to in the context of the paper. On a high level, Web3 can be thought of as an ecosystem of applications which aims to generate trust purely through decentralised technology and mathematical primitives. We posit that one of the aims of Web3 is to produce a ``Universal Trust Machine", a machine that is able to produce trust in any ecosystem, enabling long term cooperation. Thus, eliminating the need for profit driven organisations and allowing for the creation of a ``commons" \cite{commons} where everybody is free to publish, read, react, and interact with content. 

However, as shown in section \ref{problems}, fostering cooperation in a community with the presence of bad actors is not a trivial problem. In a centralised system, it is possible to govern in an ad-hoc manner, altering rules of the system as new problems and trust issues arise. This is obviously not possible in decentralised systems since no one single party can instruct everyone how to act. Therefore, all rules of interactions among the independently acting, self-interested parties must be explicitly and clearly defined before any interactions occur. Further, these rules should reasonably incentivise cooperation and disincentivise cheating/undesirable behaviour to foster long-term cooperation.

This problem of cooperation has been studied in the field of game theory and analysing studies in this field could help motivate how to develop systems where the best course of actions for neighbours is to cooperate for mutual good. 

Plethora of research also exists on models and mathematical primitives for generating trust in decentralised systems, most notably, reputation systems have gained prominence as a way to create safe and trustable communities in decentralised networks. \cite{reputationResnick}

This survey attempts to explore such mechanisms for generating trust in Web3. In section \ref{eoc} we discuss some principles in the work of Evolution of Cooperation which help motivate how long term cooperation could come about naturally. Next, in section \ref{history} we attempt to define what decentralised networks are and how the decentralised movement came about. In section \ref{web3} we explain the motivation behind Web3 and the technologies associated with it. After this, we discuss problems one faces when designing a decentralised system which fosters long term cooperation in section \ref{problems}. In section \ref{rm}, we discuss reputation systems for decentralised systems and present some promising systems in literature and the techniques they utilise for generating trust. We proceed to discuss the limitations of reputation systems and present Distributed Ledger Technologies in section \ref{dlt} which potentially remove a lot of the discussed limitations. Finally, we conclude with a future roadmap for the construction of a Universal Trust Machine.

\section{Evolution of cooperation} \label{eoc}
The history of humanity is one filled with conflict, destruction and war. The pursuit of peaceful cooperation is more than just a hippie dream, it has attracted a great deal of research across multiple fields. We believe that the goal of Web3 and the desired ``Universal Trust Machine" is to build a digital utopia where such peaceful cooperation can occur and persist over a long-term time period.

One of the foundational works investigating how cooperation can emerge and persist without a third party is ``The Evolution of Cooperation", a 1984 book written by political scientist Robert Axelrod which expanded upon the highly influential paper he co-authored with evolutionary biologist W.D. Hamilton \cite{axelrod1981}. The book's central question is ``Under what conditions will cooperation emerge in a world of egoists without central authority?".

\begin{figure*}[t]
    \centering
    \includegraphics[width=0.7\textwidth]{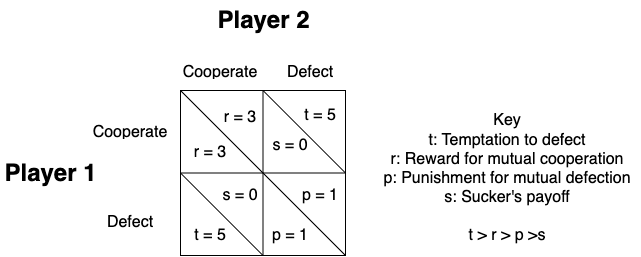}
    \caption{A typical payoff matrix of a 2 player prisoner's dilemma \cite{pd}}
    \label{fig:pd}
\end{figure*}

Axelrod held two computer simulation tournaments where multiple strategies for playing an iterated two-player Prisoner's Dilemma game were solicited from professionals across multiple disciplines. The Prisoner's Dilemma is a popular game analyzed in game theory where two rational agents are faced with a dilemma, they are arrested by the police and have to individually decide to either cooperate with the police or stay silent. The dilemma was originally framed by Merrill Flood and Melvin Dresher in 1950. A key requirement of the game is that: $t > r > p > s$ and $2 \times r > t$ where $t$, $r$, $p$ and $s$ represent payoffs for the different outcomes of the game. If both players choose to stay silent i.e. they cooperate with each other, they are each awarded $r$, on the other hand if both players defect, they are each awarded $s$. If one player stays silent while the other defects, the player who defects is rewarded $t$ while the player who chose to stay silent is paid $s$. Fig. \ref{fig:pd} demonstrates this payoff matrix visually. Hence, although the decision to collectively stay silent is overall the most optimal, individually, the best decision is to defect. 

Further, in an iterated Prisoner's Dilemma game there is a probability $w$ that two players will interact in the next round. \cite{eoc}
Contestants who submitted algorithms to play the tournament accrued points in each round according to the shown payoff matrix by playing against other strategies. The tournament consisted of five iterated prisoner's dilemma games in total with each game consisting of 200 rounds each.

The Darwinian theory of evolution would suggest that the most selfish strategy would perform the best and while indeed, in a single iteration defecting is always the best strategy, in the iterated Prisoner's Dilemma the strategy that ended up performing the best in both rounds was a simple ``Tit For Tat" strategy. As the name suggests, this strategy was based on the concept of direct reciprocity, the next move of an agent following the strategy is determined by the last move of the opposing agent, if the opposing agent cooperated, the agent following Tit For Tat would cooperate too and vice versa.

Based on the results of the tournament, Alexrod identified four characteristics that he believed led Tit For Tat to perform the best of all strategies:
\begin{enumerate}
    \item \textbf{Niceness} \\
    By being nice, Tit For Tat can benefit from long term mutual cooperation with other strategies that are also nice. However, it is important to note that niceness alone would lead to exploitation from other strategies who are not nice
    \item \textbf{Forgiveness} \\
    Strategies that are not forgiving are doomed to be locked into mutual destruction after a single defection from an opponent. Tit For Tat allows an opponent to start cooperating again after defecting initially which makes it forgiving
    \item \textbf{Retaliation} \\
    As pointed out earlier, niceness alone leads to exploitation by uncooperative strategies. By retaliating when the other strategy doesn't cooperate as expected, Tit For Tat avoids being exploited by such strategies
    \item \textbf{Certainty} \\
    By being easy to understand, Tit For Tat makes it easy for other strategies to understand what it's doing thus allowing them to come to a mutually beneficial strategy much faster
\end{enumerate}

Axelrod's analysis thus provides an interesting set of prescriptions for designing strategies for nodes on a decentralised network. Keeping in mind that not all interactions needs to be zero-sum and it may be possible for all cooperating parties to benefit on the long term by cooperating and not being the first to defect seem to work as good principles which suggest that cooperation could indeed organically grow in a pool of egoistic nodes. However, being too nice also has its downsides and any effective strategy should be quick to retaliate to prevent exploitation. Finally, keeping it simple seems to be effective advice otherwise the strategy might risk confusing potentially cooperative neighbours. 

Further, there are lessons for designers of Web3 applications, the most important being having a large ``shadow of the future", i.e. a sufficiently large $w$ which guarantees that nodes interact with each other more durably and frequently so they have time to develop a mutually cooperative strategy and since they are more likely to defect it the probability of meeting a node again is low.
This can be done in many ways including using spatiotemporal structures e.g. clustering of small groups in space \cite{pd}

However, there are limitations to Axelrod's results when considering a strategy to use as a ``Universal Trust Machine":
\begin{enumerate}
    \item \textbf{Assumptions are too simplified} \\
    Not all real word interactions are as simple as an Iterated Prisoner's Dilemma game. Often participants can communicate with each other and hence collaboration through other means may be a better strategy. Further, it may not be possible for real-world participants to necessarily perceive credible threat, or respond to it rapidly and accurately. Finally, interactions between peers are often one-time transactions, in this case, when dealing with a peer using a Tit For Tat policy, there is no incentive to behave in a trustworthy manner.
    \item \textbf{Results may not hold in some populations} \\
    In his 2000 paper "Twenty Years on: The Evolution of Cooperation Revisited", Hoffman \cite{hoffman} showed that Axelrod's tournament was sensitive to the initial population composition and the potential for strategies to make mistakes. Under different initial compositions and assumptions, other strategies were shown to perform better than Tit For Tat.
    \item \textbf{Does not consider indirect reciprocity} \\
    While direct reciprocity is a powerful mechanism, it relies on repeated encounters between individuals. However this is too simplifying an assumption to model human interactions where exchanges are often asymmetric and fleeting. Indirect Reciprocity is more representative of real human exchanges where we help people even if they've never directly helped us before based on some indirect exchange and a desire to increase our reputation in society. \cite{5rules} For example, a large-scale experiment on the prevention of blackouts found that permanent house owners (as opposed to temporary renters) and people residing in apartments were more likely participate in a demand response program to prevent blackouts when others would know their behavior and identity. This is because they were more likely to consider indirect costs and benefits to their reputation since they are more likely to have future interactions with others in the living area. \cite{houseOwnersCooperate}
\end{enumerate}

\section{Decentralisation and Decentralised Networks} \label{history}
Before considering more contemporary solutions to the problem of enabling long-term cooperation, it is important to clarify what it means to create ``decentralised" trust, what the aims of the Web3 movement are and to identify the main problems that a Universal Trust Machine should solve in order to be considered successful. In this section we discuss what decentralisation is, then in the following two sections we proceed with a similar discussion of Web3 and the inherent problems in creating decentralised trust.

Decentralisation is not a novel concept and has been prevalent in research even outside the sciences. In the social sciences, it boasts a 200 year history and has been a popular concept across multiple disciplines. Examples include concepts such as subsidiarity, democracy, liberty and equality in political science, systems theory and self determination in management and decision science, fiscal decentralisation in economics. \cite{decentralizedAI}

In technology, the concepts of technological decentralisation have been evolving for over half a century. \cite{decentralizedAI} A popular example of a decentralised IT movement is the open source software movement which represents a radical retake on copyright law and involves developing and sharing software in a decentralised and collaborative way, relying on peer review and community production. 

The importance and the success of this movement is demonstrated by the domination of multiple areas of software by open source projects. Popular examples are the the open source Apache projects which dominates the market of server software over commercial alternatives from Microsoft, Sun etc and the Linux operating system which has seen popular use being embedded in a range of devices from mobile phones, recording devices to large scale servers in data centers. \cite{openSourceApps}

\begin{figure*}[t]
    \centering
    \captionsetup{format=myformat}
    \includegraphics[width=0.7\textwidth]{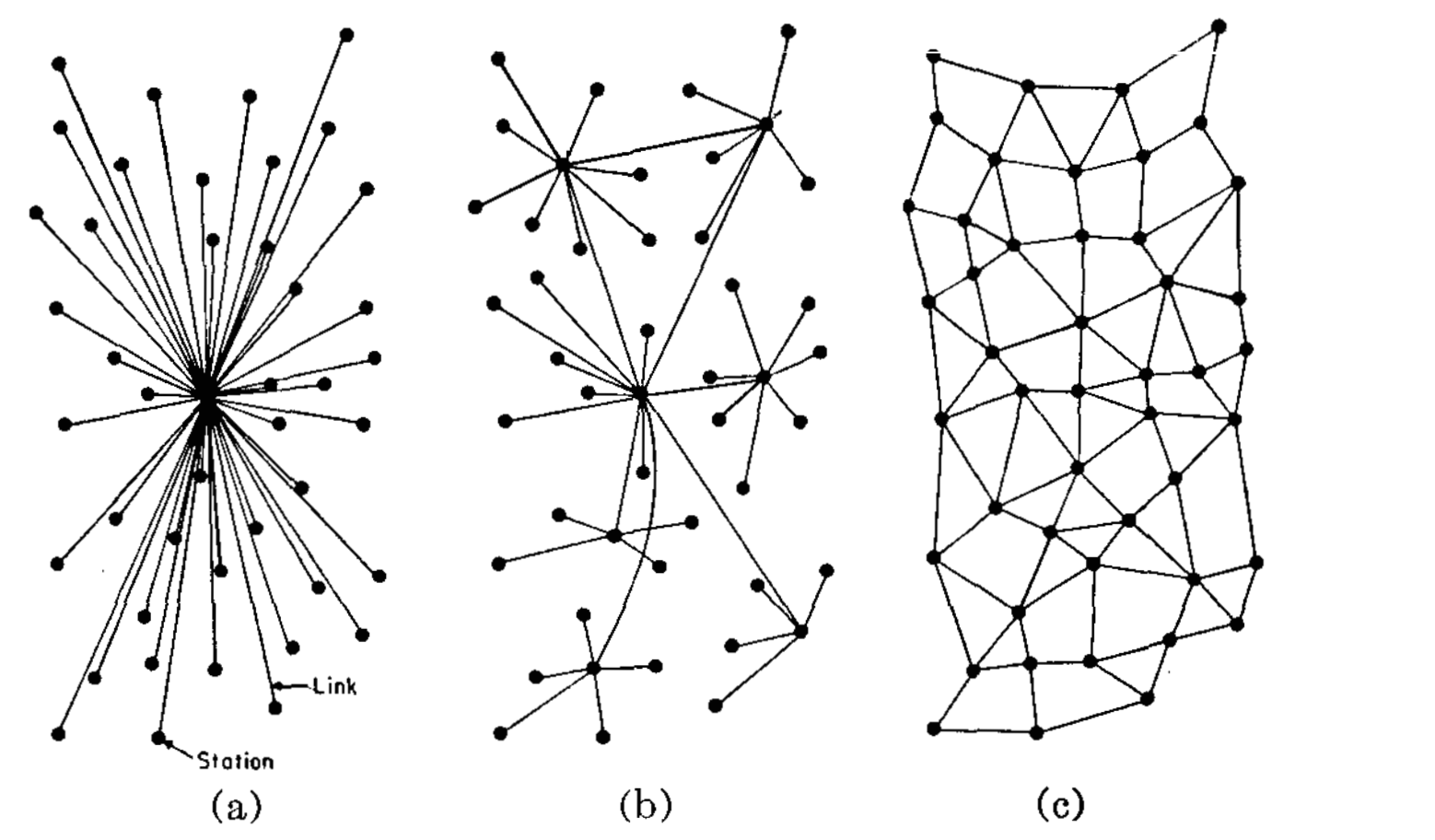}
    \caption{a) Centralised network b) Decentralised network c) Distributed network\newline
    Cardinal architectural insight from Baran's 1964 paper \cite{baran}}
    \label{fig:decen}
\end{figure*}

The concept of a ``decentralised network" was first coined by Paul Baran, one of the inventors of packet switching. In general, networks can be classified as two components, "star" or centralized and "grid"/"mesh" or distributed. In a star/centralized network, all nodes are connected to a single node, hence, each participant needs to go through a central component to interact with each other. While in a distributed network on the other hand, there is no such central node and each node can communicate with each other without going through a centralised point. In practice, a combination of these components is used to form a network, Baran called such a mixed network "decentralised" because there was no single, central point of failure. \cite{baran} Fig. \ref{fig:decen} demonstrates these networks visually.

In contemporary modern literature, the term \textit{decentralised network} is used to refer to networks where the technology, content and infrastructure on the network is controlled by participants and contributors rather than large central platforms. This control is manifested in various ways, such as participants controlling parts of the infrastructure like servers and routers, collaborators owning data in their own private data silos which is queried by the network during discovery, participants possessing the autonomy to decide the operational details of the network, what content needs to be publicised and what needs to be deleted etc. \cite{Korpal2022} In this context, a popular example of a centralised network would be Twitter which owns all the content that users publish on it, while an example of a decentralised network is Tribler, a peer to peer file sharing system which improves upon the BitTorrent protocol which enables users to share content with keyword search and boasts a reputation-management system to encourage collaboration. \cite{tribler}

Over the past decade, decentralised networks have received a reinvigorated interest due to the emergence in popularity of cryptocurrencies such as Bitcoin and Ethereum. In his whitepaper proposing Bitcoin, Satoshi Nakamoto proposed a novel decentralised peer-to-peer network protocol which facilitates an electronic payment system. \cite{satoshi} The popularity of these cryptocurrencies has also resulted in explosion of blockchain and decentralised technologies and projects. Proponents and developers of these technologies wish to see a shift in the publishing and discovery of content and information over the Internet away from a few profit-driven Big Tech corporations and into the hands of the users who generate them, guaranteeing the privacy of their data and also ensuring that everyone has a fair and equal voice.

For example, ``78 days", a collaborative project between the Starling Lab and Reuters uses decentralised ledgers to preserve historical data important to humanity. The goal of the project is to curb misinformation. It achieves so by ensuring the integrity and authenticity of the information as it captured and stored using a system called Content Authenticity Initiative. It also uses a storage system built on blockchain called Filecoin that requires data providers to prove that they are holding the authentic data and not a tampered version. Most importantly it ensures that the contributors of the information have a way to maintain their creation of the content through the records stored with the data. \cite{78days}

\section{Web3 - Decentralised Web Platforms} \label{web3}
The term ``Web2.0" was first coined by Tim O'Reilly in 2007 to describe an Internet where platforms enabled users to publish, consume and interact with content, and with each other. \cite{oreilly} It was supposed to expand upon the first iteration of the Internet or ``Web1.0" which largely consisted of static pages meant only to display information. So while ``Web1.0" was "the read web", ``Web2.0" aimed to be the ``the read-write web" (coined by Richard McManus in 2003). 

Critics of Web2.0, such as the inventor of the World Wide Web, Tim Berners-Lee feel that Web2.0 failed to achieve the vision of the Internet as a secure, decentralised exchange of public and private data, with users' data being increasingly stored in corporate data silos. Instead, to guarantee security of their data, they want users to own their own data. \cite{solid}

The term ``Web3.0" was coined by Polkadot and Ethereum co-founder Gavin Wood in 2014, he used it to describe an Internet that is decentralised, open and transparent. \cite{wood}

The current Web3 movement aims to transform the platform oriented Web2.0 Internet into a decentralised web ecosystem which: 1) avoids monopoly of content discovery and propagation by large centralised actors 2) prevents the spread of misinformation and fake news 3) provides its users the ability to create, exchange and react to information in a secure, private and free manner 4) supports immersive web development \cite{decentralizedAI}

Liu et al \cite{liu} define Web3 as a movement which agnostic of any specific overarching applications or underlying infrastructures will usher in ``an era of computing where the critical computing of applications is verifiable", that is, an application that conforms to the idea of Web3 is one where all stakeholders are able to verify the execution of the application based on predetermined terms without the presence of an intermediary.

Packy McCormick defines Web3 as ``the Internet owned by the builders and users, orchestrated with tokens". \cite{garon} Defining Web3 with its key property being user ownership is a common approach taken by a majority of research papers on the topic. Hence, Web3 is positioned as the ``read, write, own" web. While Web2.0 was a frontend revolution that allowed users to create and interact with created content online, Web3 is instead a backend revolution which aims to change how the created content is stored. Instead of keeping data on centralised data silos, Web3 aims to provide data storage to users in a distributed manner in a way that users can own and monetise the content they created. Thus, it aims for the disintermediation of existing parties such as large big tech companies in data governance. \cite{interoperability}

Finally, in addition to personal ownership of data, many Web3 proponents also believe in the concept of \textit{``Self-Sovereign Identity"} i.e. that identity holders on the Internet should also be owners of their identities. Centralised identity solutions require holding many plastic cards and username/passwords, leaving individuals with little control of their identity and prone to privacy theft. A Web3 with \textit{Self-Sovereign Identity} would allow users to have a persistent, transparent identity which they can control fully e.g. decide which platforms have access to their identity and what information they can view. \cite{identity}
\section{Threats to long term cooperation} \label{problems}
In order to enable the dream of Web3, it is fundamental to be able to create a commons with communities of users interacting with each other through decentralised networks, free to read, publish and interact with content. However, two broad classes of threats make creating long term cooperation in decentralised networks a non-trivial task: Social and Infrastructural threats. In the following sections we briefly cover these threats and establish why they pose a problem to cooperation.
\subsection{Infrastructural Threats}
In section \ref{intro}, we motivated why trust is fundamental to achieving cooperation inside communities. Since in a Web3 application based on a decentralised network there are no third parties for enforcing trust, before using a service to cooperate with other nodes in the network, users look for assurance that the other party is trustable. This is especially true for applications that depend on blockchain technology due to the immutable nature of transactions making it incredibly hard to punish bad actors. \cite{blockchainTrust} Therefore, in addition to social problems, there are also several infrastructural problems stemming from the presence of bad actors who wish to abuse the trust of their neighbours for their own benefits makes the problem of achieving long term cooperation in a decentralised network a non-trivial task. A system that will be able to achieve the stated dreams of Web3 should be effectively able to tackle these problems, below is a brief description of a few of these problems:

\subsubsection{Sybil Attack}
In a distributed network, if an entity can control a large number of nodes and hence obtain a large number of node identifiers, they can use this dominance of identities to control the network and undermine the mechanisms of the network which results in a network with less robustness and freedom. Such an attack is often referred to in literature as a \textit{Sybil Attack}, where a \textit{Sybil} is the fake identity of an entity. \cite{SybilTax}

The Sybil Attack was first mentioned by Doucer in \cite{SybilDoucer}. In this paper, Doucer argues that only a central authority can prevent a Sybil Attack under realistic assumptions of resource distribution and coordination.

While the intuitive solution to making a network robust against a Sybil Attack seems to be to make it expensive to create new identities in the network, doing so increases the social cost of the network by making it hard for new users to join it.

In the context of reputation systems in decentralised networks, a colluding group of malicious nodes could also increase the reputation of its nodes by itself and hence threaten the integrity of the network.

\subsubsection{Free riding} 
In order to encourage successful long term cooperation, it is important that enough peers are providing sufficient resources for the system to become large and truly useful. In the absence of a third party monitoring each user, it is possible that some users stop contributing and only consume resources being generated by other users. \textit{Free riders} are peers that eagerly consume resources without reciprocating any in return. It is easy to see how free riders diminish the quality of service for other peers, but more importantly, by making contributing peers feel exploited they disincentivise cooperation in the system and thus threaten the existence of the whole system, especially systems that are predicated on the foundation of sharing. 

However, in the context of a decentralised network the most important problem created by free riding is that if only a few users are providing resources, they end up acting as centralised servers, this threatens the security of the network and defeats the very goal of the Web3 application.

Gnutella is a popular peer-to-peer file sharing platform which allows users private access to information. In their paper ``Free Riding on Gnutella", Eytan Adar and Bernardo A. Huberman \cite{Huberman} showed that 70\% of Gnutella users were not sharing any files and nearly 50\% of responses for file discovery were being returned by the top 1\% of sharing hosts. 

Similarly, Locher at al \cite{bittorent} were able to create ``BitThief", a free riding BitTorrent agent that was able to achieve high download rates even without seeding any data in return. They were also able to demonstrate that sharing communities which originally intended to promote cooperation among peers ultimate provide many incentives to cheat.

\subsubsection{Pollution Attack}
In 2005, Liang et al \cite{liang} showed that it was possible for an attacker in a decentralised network to corrupt certain targeted content, rendering it unusable and then making it available to the network in a large quantity. Since users on the network are unable to distinguish between the polluted and the original content through content discovery alone, users download the polluted content and further share it with other peers, resulting in the polluted content spreading through the network.

In their analysis of the FastTrack peer to peer sharing system, it was found that as many as 50\%-80\% of copies of popular content were polluted.
\subsubsection{Index Poisoning}
Often resource sharing in decentralised networks is conducted through indices, which allow users to conveniently discover the location of their desired content. Depending on the architecture of the system, the index could be distributed over a fraction of the file sharing nodes (as in FastTrack) or over all the nodes. 

In an Index Poisoning attack an attacker inserts bogus records into the index, for example, by inserting random identifiers that do not correspond to any address into the index. This way, when a user attempts to download a file they are unable to locate its content, leading to them finally abandoning the search. \cite{indexPoisoning}

While the Pollution attack described earlier requires the attacker to obtain high-bandwidth to make sufficient versions of the corrupted copies available in the network, the Index Poisoning attack is easier in that it requires less resources to pull off.

\subsubsection{Slandering}

Under Sybil Attack, we discussed that it may be possible for a colluding group of malicious nodes to do \textit{self promotion} to increment their own reputation in a reputation based decentralised system. On the other hand, it may also be possible for a group to coordinate to reduce the reputation of a victim, such an attak is called \textit{slandering} \cite{simFrame}

\subsubsection{White Washing}
Nodes that have accrued a bad reputation by acting in an undesired manner can "clean" a bad reputation through \textit{white washing} to avoid the negative effects of having a bad reputation \cite{simFrame}

\subsubsection{Denial of service}
Cooperating nodes can work to block the functioning of a decentralised system, preventing other peers from utilizing its services
\subsection{Social Threats}
As seen by recent events, the rise of populist movements stands to be the biggest threat to the state of democracy worldwide. Many observers, especially journalists have suggested that the rise and spread of these movements has been massively aided through social media. \cite{populism} While social media can be a powerful tool for spreading information, when left unregulated, it can also lead to multiple social issues which greatly threaten long term cooperation. Some of these issues are:
\subsubsection{Echo Chambers and Polarisation} 
``Echo Chambers" are used to describe the mechanism by which people on sociotechnical platforms are exposed to large or exclusively pro-attitudinal communication. Such grouping of like minded people on social networks (`homophily') is believed to arise from preferential connection to like minded individuals when creating/breaking bonds and also from peer influence which results in connected individuals growing more similar. \cite{polarisationClimateChange} The presence of an Echo Chamber could support populist messages that support rejection of expertise and reasoned debate among different views and lead to the emphasis of popularity of people or ideas over substance of their views. Therefore, Echo Chambers can lead to an insulation of users from the truth and even more perniciously, to be exposed to fake news.
\\
In their study on Echo Chambers in the context of COVID-19 discussions on Twitter, Jiang et al \cite{echoChambersCovid} found strong evidence of political echo chambers on the topic on both ends of the political spectrum, but particularly so in the right-winged community. They found that tweets by right leaning users were almost exclusively retweeted by users who were also right leaning. Further, from random walk simulations, it was found that information in right leaning bubbles rarely travelled out of that bubble, forming a ``small, yet intense political bubble".
\\
In another study on Climate Change discussions on Twitter, Williams et al ``found a high degree of polarisation in attitudes, consistent with self selection bias" \cite{polarisationClimateChange}
\\
Studies have suggested that echo chambers could lead to polarisation of users and thus to users retreating into like-minded networks \cite{norris2019cultural}, which creates segmentation in networks and thus poses a large challenge to long term cooperation.

\subsubsection{Inequality and Social Divide} 
While the idea of a digital democracy is appealing, it is hampered by findings of socioeconomic inequality which prevent usage of the platforms by certain stratas of society. Beyond inability to access platforms, it is possible that members of society lack the skills to express their views or consume information that is being shared by other members. \cite{socialMediaInequality}
\\
A lack of participation by different members of society could lead to the propagation of biased views or misinformation against the underrepresented members. Thus, it constitutes a credible threat to long term cooperation.
\\
However, diffusion theories predict inequality at the outset of any innovation which is narrowed as time progresses and adoption rate spreads.

\section{Reputation Systems} \label{rm}

As motivated at the end of section \ref{eoc}, instead of only relying on direct reciprocation in decentralised systems, we can allow users that help each other out to establish a good reputation which can be used to reward them in some other way. After all, this is more representative of real social interactions, while we are interested in how people interact with us, we are also interested in the actions of others which we learn about from social channels such as gossip. In taking actions, we don't only take into account our direct experiences but also experiences we've learnt about from indirect sources. Similarly, when choosing to assist someone we also consider how it affects our reputation in society.

Although animals possess simple mechanisms for indirect reciprocity, only humans engage in complex reputation systems. \cite{5rules} This seems to be because such systems require a substantive cognitive load, not only does it require a memory of all transactions but also requires the ability to monitor the dynamically changing social network of the group. Hence, the strategies required to succeed in indirect reciprocity are also understandably a lot more complex than the simple Tit For Tat strategy that succeeds in direct reciprocity.

In their paper on reputation systems, Resnick et al \cite{reputationResnick} define a reputation system as one that ``collects, distributed and feedback about participants' past behavior ... these systems help people decide whom to trust, encourage trustworthy behavior, and deter participation by those who are unskilled or dishonest."

As mentioned before, users on decentralised networks look for some form of assurance that their transactions on the network will be successful. The reputation of a user in reputation systems serves as a ``shadow of the future" to each transaction, creating an expectation for what a user can expect when dealing with another user.

Consider the example of one of the first reputation system in eBay, the ``Feedback Forum": after a transaction is completed, a buyer or seller can rate each other (1, 0 or -1) and leave comments. A participant in eBay accumulates such points over time which are displayed next to their screen name. A buyer can view a seller's points and comments left by other users to create a ``shadow of the future" into the transaction they can expect to have if they buy an item from the seller. Many other online forums and marketplaces such as Amazon and Stack Overflow rely on similar reputation systems.

According to Resnick, a reputation system must meet three challenges: \cite{resnickEbay}
\begin{enumerate}
    \item Provide information that should allow users to distinguish between trustworthy and non trustworthy users,
    \item Encourage users to be trustworthy, and
    \item Discourage participation from users who aren't
\end{enumerate}

In addition to the above, a successful reputation system should also be able to avoid issues mentioned in \ref{problems}

\begin{table*}[t]
\centering
\begin{tabular}{|c|c|c|c|c|}
\hline
\textbf{Year} & \textbf{Reputation Mechanism} & \textbf{Trust Function} & \textbf{Decentralised} & \textbf{Sybil Tolerant} \\
\hline
1999 & PageRank     &        \makecell{Trustworthiness of a node is determined by counting \\ the number and quality of links to it. \\ $PageRank$ = $\sum$ $\frac{PageRank\ of\ Inbound\ Link}{Number\ of\ Outgoing\ Links\ on\ that\ Page}$ \\}               & $\times $             &           $\times $       \\
\hline
2003 & EigenTrust           &     \makecell{Distributed PageRank, allows for calculation of a global trust value \\ which reflects the experience of all peers in the network. \\ $t_{ik} = \sum_j c_{ij} c_{jk}$}            &      $\checkmark $         &     $\times $             \\
\hline
2004 & PeerTrust           &       \makecell{To determine trustworthiness, in addition to the feedback from other peers, \\ also takes into consideration other factors such as credibility of feedback, \\ the transaction and community context.\\ $T(u) = \alpha \times \sum_{i=1}^{I(u)} S(u, i) \times Cr(p(u, i)) \times TF(u, i) + \beta \times CF(u)$}          &      $\checkmark $         &     $\times $             \\
\hline
2009 & BarterCast           &       \makecell{Aims to prevent lazy freeriding in p2p filesharing applications.\\ Local representation of network is created at each node, \\ reputation of peer is then calculated using a maxflow algorithm \\ $R_i(j) = \frac{\arctan(maxflow(j,i) - maxflow(i,j))}{\pi/2}$}          &      $\checkmark $         &     $\times $             \\
\hline
2012 & WikiTrust            &    \makecell{Trustworthiness of user is calculated by measuring meaningfulness of their contribution \\ This meaningfulness is calculating using an edit distance function.\\ $q(b|a,c) = \frac{d(a,c) - d(b,c)}{d(a,b)}$}             &    $\times $            &        $\times $          \\
\hline
2015 & HonestPeer           &       \makecell{Enhanced EigenTrust, reduces dependency of original algorithm on a set \\ of pre-trusted peers by selecting a group of highly selected peers dynamically\\ $t^{k+1}_i = \begin{cases}
                                   a \times p_i + (1 - a) \times \sum_{x=1}^{x=n} c_{xi}t_x^k & \text{if $h  \in P$} \\
                                   (1 - a) \times p_i + a \times \sum_{x=1}^{x=n} c_{xi}t_x^k & \text{if $h  \notin P$} 
    \end{cases}$}          &      $\checkmark $         &     $\times $            \\
\hline
2022 & MeritRank            &     \makecell{Provides a set of decay based constraints which help to provide \\ Sybil resistance to otherwise Sybil prone strategies \\ Where, a trust graph modelled using constraints satisfies: \\ $\lim_{|S|\to\infty}$ $\frac{w^+(\sigma_s)}{w^-(\sigma_s)}$ $\leq c$ \\
}            &       $\checkmark $        &     $\checkmark $          \\
\hline
\end{tabular}
\caption{Overview of mentioned reputation systems}
\label{table:1}
\end{table*}

The following are a few notable reputation systems which attempt to accomplish the objectives stated above:

\subsection{PageRank}
One of the most widely known reputation systems in the world is Google's PageRank. PageRank determines a rough estimate of the relative importance of a website by computing a ranking for every web page. The underlying assumption of PageRank is that a website that is more important is more likely to receive links from other websites than a website that is less important. PageRank is an interesting example of a Reputation Mechanism since while it may not be the exclusive algorithm used by Google, it has inspired many other reputation algorithms.

The calculation of PageRank of a website can be simplified to the below equation: \\
\begin{equation}
\sum \frac{PageRank\ of\ Inbound\ Link}{Number\ of\ Outgoing\ Links\ on\ that\ Page} \\
\end{equation}

Hence, if a website $a$ with a high PageRank has a link to another website $b$, website $b$ will receive a large boost to its PageRank. However, the contribution of $a$'s PageRank to $b$'s PageRank will be reduced if $a$ has a lot of outgoing links, this is ensured by dividing the contribution of each inbound link by the number of outgoing links on that page. 

Through this simple idea, Google was able to very successfully rank websites in terms of relevance. The idea was so revolutionary that PageRank is still used in Google today (along with 200 other more complex algorithms). \cite{PageRank} However, PageRank relies on a Trusted Oracle model which requires a centralised service, dependency on such an oracle to provide reputation introduces points of failure and does not scale well. Further, the original version of PageRank is susceptible to Sybil attacks. \cite{SybilPageRank}
\subsection{WikiTrust}
\textit{WikiTrust} \cite{wikitrust} is the reputation system used for one of the largest collaborative applications known to mankind: the writing of articles on Wikipedia. It is a content-driven reputation system, that is, it relies on automated analysis of the content generated by the user and the collaboration process to derive the reputation of the user, rather than explicit feedback provided by users on other users. It is possible to use such a reputation system since the applications it's catered for is entirely content driven.

The goals of \textit{WikiTrust} are to incentivise lasting, meaningful contributions from users, help increase the quality of content being produced, spot vandals and to offer users an indicator of the quality of the content they are consuming. To achieve these goals, WikiTrust maintains different reputations for users and the content they create.

If a user makes a contribution that is meaningful and its content is preserved in future edits, they gain reputation, on the other hand, if their contributions are wholly or partially undone by future edits, then they lose reputation. Content starts with no reputation, if they are revised by users with high-reputation, it gains reputation. On the other hand, if the text is disturbed by too many edits, indicating that the content may not be trustworthy, it loses reputation.

In order to estimate how much each contribution is preserved or removed as required for the above, WikiTrust relies on an edit distance function $d(r, r')$ which is computed based on how many words have deleted, inserted, replaced and displaced from the edit that led from $r$ to $r'$. Relying on such a distance functions allows the reputation system to be language independent. Finally, the value of an edit is calculated using the function: \\
\begin{equation}
q(b|a,c) = \frac{d(a,c) - d(b,c)}{d(a,b)} \\
\end{equation}
Where $b$ is the edit being evaluated, $a$ is the revision before the edit and $c$ is the revision after it. $q(b|a,c)$ outputs a value between -1 and +1;  it is equal to -1 if $a=c$ and hence implying that $b$ was entirely reverted, on the other hand, it is equal to +1 if the change from a to b was entirely preserved. However, a limitation of this approach is that since it requires subsequent revisions, it is unable to judge newly created revisions. 

WikiTrust only considers not negative reputation values, new users are assigned a reputation very close to 0, this ensures that vandals cannot white wash themselves since their new identities would have a similar reputation to their vandal identity. Also, due to the content driven nature of the system, creating Sybils is harder than in a system where identities can simply be used to promote each other.

\subsection{EigenTrust}
While the reputation systems listed so far possess many interesting properties, both of them require a centralised ``oracle" which acts as an intermediary for all nodes in the network, aggregating and providing trust values when a node requests them. Such an oracle is antithetical to the design of a decentralised application.

EigenTrust \cite{eigenTrust} could be described as a distributed version of PageRank. The algorithm allows the calculation of a unique \textit{global trust value} $\overrightarrow{t}$ for each peer $i$ in the network which reflects the experience of all peers in the network with peer $i$. $\overrightarrow{t}$ provides $i$ a trust value for each peer in the network which it can refer to in order to establish how much it can trust another peer, ensuring it only conduct transactions with trustworthy peers. Further, it also has mechanisms to ensure that a malicious group of cooperating peers cannot lie for their own benefit.

Similar to eBay's reputation system, the system requires each peer to rate another peer after it conducts a transaction with them. This results in the creation of a local trust value $s$ for each peer where $s_{ij}$ reflects how much $i$ trusts peer $j$ based on its transactions with them. It is suggested that one way of calculating $s_{ij}$ is using:
\begin{equation}
s_{ij} = sat(i,j) - unsat(i,j)\\
\end{equation}

Where $sat(i,j)$ and $unsat(i,j)$ represent the number of satisfactory and unsatisfactory transactions that $i$ had with $j$ respectively. These localised trust values $s_{ij}$ are further normalised to produce $c_{ij}$ to ensure that the trust values are between 0 and 1, to ensure that malicious peers can't simply assign arbitrarily high local trust values to other malicious peers and low values to other peers, abusing the system. 

The local trust values are then aggregated in each peer to produce $t$ as below:
\begin{equation}
t_{ik} = \sum_j c_{ij} c_{jk}\\
\end{equation}

This is equivalent to $i$ asking its acquaintances how much they trust their acquaintances. However, so far $t$ only reflects the experience of $i$ and its acquaintances. This process needs to be repeated again in order to reflect $i$'s acquaintances' acquaintances experience and so on. The authors of the paper are able to prove that the final trust vector $\overrightarrow{t_i}$ will converge to the same vector $\overrightarrow{t}$for every peer $i$ in the network, namely the left principal eigenvector of the matrix $[c_{ij}]$

$\overrightarrow{t}$ is calculated in a distributed manner. The authors prove that in a network with where the number of active peers are small, this can be done relatively efficiently since each peer has limited transactions. In a network with a large number of peers, the algorithm can be performed efficiently by limited the local number of local trust values $c_{ij}$ t)at each peer can report. Further, a decay factor $a$ can be used to reduced the influence of malicious cooperating peers ensuring that the same group of nodes vouching for each other is not as significant in the resulting trust values as a diverse group.

The main challenge facing the design of distributed reputation systems is the aggregation of local trust values into global trust values. EigenTrust's translated the notion of \textit{transitive trust} where if a peer trusts another peer it also trust its trusted network into a distributed algorithm which can run efficiently without congesting the network.

While EigenTrust is a powerful method, it requires the presence of a prior notion of trust i.e. a group of peers that are known to be trustworthy. The authors suggest that this could be the first few peers that join the network since the designers and early users of a P2P network are less likely to want to cheat in a network that they helped create. However, this assumption is a significant disadvantage of this Reputation Mechanism.
\subsection{HonestPeer}
EigenTrust's assumption of a static group of trusted peers marginalises other peers, resulting in them being ranked much lower despite them potentially being honest. It also leads to potential poisoning vulnerabilities since if a trusted peer downloads a poisoned file from a malicious peer, it could result in the network also downloading the file. Finally, relying on a selected group of nodes comes with a lot of the same problems as relying on a single central entity.

HonestPeer \cite{honestPeer} is an enhanced version of EigenTrust which tackles this problem by giving peers with a high reputation value a role in calculating the global reputation of other peers. Hence, instead of solely relying on a static group of peers, the algorithm selects a group of highly trusted peers dynamically, making it more robust and less centralised. While several improvements have been suggested to the original EigenTrust algorithm, HonestPeer is notable because it is able to reduce the algorithm's dependency on pre-trusted peers without sacrificing the simplicity of the algorithm, an important requirement for effective cooperation strategies as seen in section \ref{eoc}

The implementation follows the same approach of calculating trust values for each peer as EigenTrust. The trust values calculated in each run are used to find a honest peer $h$ for each node, where:

\begin{equation}
t^{k}_h = \max_i{(t^{k}_i)}
\end{equation}

This honest peer, $h$ is then used in the calculating the proliferation parameter $a$, where:

\begin{equation}
a = \begin{cases}
                                   t^{k}_h & \text{if $t^{k}_h  > 0.5$} \\
                                   1 - t^{k}_h & \text{if $t^{k}_h  \leq 0.5$} \\
\end{cases}
\end{equation}

Based on this, the current reputation of peer $i$ is then calculated as:

\begin{equation}
    t^{k+1}_i = \begin{cases}
                                   a \times p_i + (1 - a) \times \sum_{x=1}^{x=n} c_{xi}t_x^k & \text{if $h  \in P$} \\
                                   (1 - a) \times p_i + a \times \sum_{x=1}^{x=n} c_{xi}t_x^k & \text{if $h  \notin P$} 
    \end{cases}
\end{equation}

Where $P$ is the group of pre-trusted peers. As a result of this modification, the influence of the pre-trusted peers is high if $h$ is a part of them, otherwise their effect on the reputation is marginalised. Through simulation in a p2p file sharing network, the paper's authors were able to demonstrate that HonestPeer reduced the percentage of invalid files and increased the success rate of good files downloaded. Further, by allowing $h$ to be dynamically replaceable, the algorithm ends up more scalable too, which was also demonstrated in simulation.

\subsection{BarterCast}
BarterCast \cite{barterCast} is a lightweight, fully distributed reputation system that aims to prevent \textit{lazy freeriding} in P2P file-sharing systems. Lazy freeriding in this context is differentiated from \textit{die-hard freeriding} which consists of nodes employing sophisticated methods to subvert the reputation system. The authors of the paper argue that only a small fraction of users in any application actually use sophisticated measures and hence, it is prudent to create an efficient reputation system that can at least prevent lazy freeriding.

To establish the local subjective reputation of peer $j$ at node $i$, $i$ builds a local representation of the network which is used to derive the reputation of $j$ based on: a) direct experience of $i$ with $j$ b) information about $j$ that $i$ receives from other peers. 
The local representation of the network is constructed using a directed graph where edges to other nodes exist if there is a direct interaction between two nodes. After the local graph has been constructed, a \textit{maxflow} algorithm is run on the graph where given a graph $G$, the capacity $c(i,j)$ represents the \textit{total number of bytes} transferred from one peer to another and hence the flow, $f(i,j)$ is a measure of ``indirect service" received by $i$ from $j$ in the network. The subjective reputation $R_i(j)$ of peer $j$ at node $i$ is then calculated as:

\begin{equation}
    R_i(j) = \frac{\arctan(maxflow(j,i) - maxflow(i,j))}{\pi/2}
\end{equation}

$R_i(j)$ calculated this way has a value between -1 and 1. The \textit{arctan} function is used in the equation to ensure that even modest contributions by new peers can result in a significant reputation. In order to ensure practical efficiency, the implementation only regards paths with a maximum length of two. The paper's authors claim this to be a reasonable assumption given the \textit{small-world effect}, where 98\% of peers exchange data directly or with a common third party.

The local representation of the network is built in two-ways: a) Using the private history of peer $i$ where in an entry $(j,up,down)$ $up$ is the number of bytes $i$ has uploaded to $j$ and $down$ the number of bytes $i$ has downloaded from $j$ and b) Using an exchange of private history between peers through messages.

The reputation calculated using BarterCast can then be used to prioritise upload bandwidth to peers with a high reputation or to rate limit peers with a low reputation.

\subsection{PeerTrust}
Like all the reputation systems mentioned above, PeerTrust \cite{peerTrust} evaluates a node's trustworthness by taking in consideration the feedback a peer has obtained from other peers. However, in addition to simply considering the feedback, PeerTrust also takes into account certain other factors:

\begin{enumerate}
    \item \textbf{Credibility of Feedback} \\ It may be possible for a peer to lie in their feedback due to malicious motives. Therefore, the credibility of a node is taken into account when deciding how much to value their feedback of another node.
    \item \textbf{Transaction Context Factor} \\ Not all transactions are equal, for example, in an application that involves transactions between users, transactions of greater economic value should influence a user's trust more than transactions of small value since otherwise a user could behave in a trustworthy manner in a lot of transactions of small value but cheat in one transaction of large value and still end up with a positive reputation. Beyond this, just because a node can provide good services in a certain context, this does not necessarily imply that they can provide comparable services in a completely different context, for example, a node that provides good information about tourism should not also automatically be trusted to provide equally good medical advice. Therefore, the context of the transaction is made a factor when calculating trust.
    \item \textbf{Community Context Factor} \\ In order to deal with community specific issues and vulnerabilities such as lack of incentive to provide feedback and collaboration of malicious peers to manipulate their trust scores, community contexts are added as a factor when calculating trust.
\end{enumerate}

Hence, the trust value $T(u)$ of a node $u$ is defined as:

\begin{equation}
    T(u) = \alpha \times \sum_{i=1}^{I(u)} S(u, i) \times Cr(p(u, i)) \times TF(u, i) + \beta \times CF(u)
\end{equation}

Where,
\begin{itemize}
    \item \textbf{$I(u)$} denotes the total number of transactions of $u$ with all other nodes in a recent time window
    \item \textbf{$p(u, i)$} denotes the other node that participated in peer $u$'s $i$th transaction
    \item \textbf{$S(u,i)$} is the normalised amount of satisfaction that $u$ received from node $p(u,i)$ in the $i$th transaction
    \item \textbf{$Cr(v)$} is the credibility of the feedback submitted by $v$
    \item \textbf{$TF(u,i)$} is the adaptive transaction context factor for node $u$'s $i$th transaction
    \item \textbf{$CF(u)$} is the adaptive community context factor for node $u$
\end{itemize}

\subsection{MeritRank}
\textit{MeritRank} \cite{meritRank} uses a merit based tokenomics model which aims to bound the benefits of Sybil attacks instead of preventing them altogether. The system is based on the assumption that peers observe and evaluate each others' contribution, similar to the reputation system used in eBay. Each peer's evaluation is stored in a personal ledger and modelled in a feedback graph where the feedback to each user is modelled as a special token value which accumulates over time. It is also assumed that each peer is able to discover the feedback graph, for example, through a gossip protocol.
MeritRank manages to achieve this Sybil tolerance by imposing the following constraints on how reputation can be gained inside the feedback graph:
\begin{enumerate}
    \item \textbf{Relative Feedback} \\ This constraint places a bound on how much feedback a single entity can provide to another entity by the degree of the entity i.e. the size of the set of its neighbours. This constraints assists in limiting a single entity from creating multiple parallel Sybils
    \item \textbf{Transitivity $\alpha$ decay} \\ This constraint limits the ability of an entity to create a serial Sybil attack by terminating random walks in the feedback graph with a probability $\alpha$
    \item \textbf{Connectivity $\beta$ decay} \\ Sybil attack edges in a feedback graph are often bridges i.e. their cut creates two separates components. This constraints introduces a punishment for a node for being in a separate component
\end{enumerate}
A trust graph modelled using these MeritRank's constraints will satisfy:
\begin{equation}
\lim_{|S|\to\infty} \frac{w^+(\sigma_s)}{w^-(\sigma_s)} \leq c \\
\end{equation}
where, $w^+(\sigma_s)$ is the profit gained by the Sybil Attack $\sigma_s$, $w^-(\sigma_s)$ is the cost of the Sybil attack, $S$ is the set of Sybils and $c$ is some constant value such that $c > 0$. Thus MeritRank is able to provide a reputation system with feedback which is Sybil tolerant.

Table \ref{table:1} summarises the reputation systems discussed so far. The survey of reputation systems above in not an exhaustive list, for a more comprehensive treatment of the subject the reader is referred to the following surveys: \cite{survey1,survey2,survey3,survey4}. However, the reputation systems stated above provide a decent summary of how indirect reciprocity can be used to create trust in a decentralised network while also tackling a lot of trust issues inherent to the decentralised setting. 

While indirect reciprocity through reputation systems is a powerful tool for tackling the infrastructural threats in \ref{problems} such as Sybil attacks, whitewashing etc., in of themselves, reputation systems cannot serve as an ``Universal Trust Machine" due to the following limitations:
\begin{enumerate}
    \item \textbf{Not Privacy Preserving} \\ While a lot of reputation systems presented so far may be confidentiality preserving, they are not privacy preserving, i.e. they do not prevent the discovery of users who contributed to a reputation rating. For example, if a node goes offline between two reputation queries, the difference in the aggregated reputation score across the two queries can reveal the user's contribution. \cite{bbdt} 
    \item \textbf{Do not provide a mechanism to carry out transactions} \\ While reputation systems are great for applications such as P2P file exchange, enabling cooperation in other domains such as e-commerce and IoT requires mechanisms beyond simple provision of trust such as the ability to carry out transactions.
    \item \textbf{Lack of Flexibility} \\ Guaranteeing trust in the real-world is often not as simple as simply identifying trustable users, centralised institutions provide bespoke functionality such as financial contracts and escrows in order to enable cooperation between users who want to perform transactions with each other. In section \ref{dlt}, we show how Smart contracts can allow for such functionality in the distributed world.
    \item \textbf{Requires all entities to remain online} \\ Most of the decentralised solutions suggested above suffer from the problem of requiring trusted entities to always be online and connectable in order to enable network discovery and furthermore, have a valid address where they can be contacted. This is too large an assumption in a lot of domains such as IoT where nodes are constantly going offline.
    \item \textbf{Do not solve social threats} \\ It is important to note that reputation systems only tackle infrastructural threats. Due to their limited scope, they are unable to tackle the social threats listed in section \ref{problems}.
\end{enumerate}

In the next section, we present Distributed Ledger Technology (DLT), a technology that offers a solution to a lot of the limitations listed above.

\section{Distributed Ledger Technology} \label{dlt}

Any technology that leverages a ledger in order to store data distributed across multiple nodes in referred to as a DLT. The recent rise in popularity of Bitcoin has led to prominence of Blockchain technology and multiple other technologies that leverage Blockchain such as Ethereum, Hyperledger, Cardano etc. However, Blockchain is not the only DLT, some other examples of DLTs include: Tangle, Hashgroup and Sidechain. \cite{dltr} presents an extensive comparison of these technologies.

At their core, DLTs are data structures that allow recording of transactions and functions for their manipulation. Generally, all DLTs are based on three well-known foundational technologies: \cite{dltr}

\begin{enumerate}
    \item \textbf{Public Key Cryptography} \\ Since DLTs operate in insecure, distributed environments, Public Key Cryptography allows for the establishment of secure digital identities and communications between nodes. The digital identities also allow for the enforcement of ownership of resources in the network and thus helps facilitate transactions.
    \item \textbf{Distributed Peer to Peer Network} \\ Operating in a distributed network allows for a highly scalable network without a single point of failure (as in a centralised network).
    \item \textbf{Consensus Mechanism} \\ The presence of such a mechanism allows for all nodes in the network to converge on a single version of global truth without a trusted intermediary.
\end{enumerate}

DLTs are similar to reputation systems in that often, their main goal is to allow interactions between users that do not trust each other without a trusted third party. \cite{ukgovdist}
By design, DLTs offer a high degree of transparency, traceability and security which allows them to offer security, privacy and trustworthiness inside a diverse set of applications. Below is a survey of four DLT technologies which helps to demonstrate the benefits provided by them:

\begin{table*}[t]
\centering
\resizebox{\textwidth}{!}{\begin{tabular}{|c|c|c|c|c|c|c|}
\hline
\textbf{Year} & \textbf{DLT} & \textbf{Ledger}  & \textbf{Low Fees} & \textbf{High Scalability} & \textbf{Smart Contracts} & \textbf{Sybil Resistance} \\
\hline
2008 & Bitcoin     &     Global Blockchain               &  $\times $              &         $\times $ & $\times $ & $\times $      \\
\hline
2014 & Ethereum           &    Global Blockchain               &  $\times $              &         $\times $ & $\checkmark $ & $\times $      \\
\hline
2018 & IOTA           &       Tangle          &  $\checkmark $              &    $\checkmark $ & $\times $ & $\times $      \\
\hline
2018 & TrustChain            &    Locally Stored Linked Blockchains    &  $\checkmark $              &    $\checkmark $ & $\times $ & $\checkmark $      \\
\hline
\end{tabular}}
\caption{Overview of mentioned DLTs}
\label{table:2}
\end{table*}

\subsection{Bitcoin}
Bitcoin \cite{satoshi} is a cryptocurrency that leverages Blockchain technology to offer a distributed and immutable ledger that stores transaction history. The core technology consists of a linked list of blocks that are connected together, with each block referencing the previous block in the chain. Transactions are continuously being appended in blocks to the chain and are visible to all participants in the network.

Bitcoin uses Proof of Work (PoW), a form of cryptographic proof as a consensus mechanism which allows a party to prove to other nodes in the network that a specific amount of computational power was expended. The nodes in the network are able to verify this expenditure with minimal effort. The purpose of using PoW as consensus mechanism is to deter manipulation of data in the ledger by imposing an infeasible energy and hardware requirement in order to do so, thus guaranteeing security of the ledger and allowing nodes in the distributed network to conduct transactions with each other.

However, multiple papers have criticised the inefficiency of PoW and Bitcoin \cite{inefficiency1, inefficiency2}, notably the requirement of wasting enormous amount of electricity through expensive mining equipment. Further, transactions on the network require large transaction fees which have to be paid to miners and a large confirmation time for transactions, making Bitcoin a poor choice for applications that require a large amount of transactions, quickly.

\subsection{Ethereum}
Similar to Blockchain, Ethereum \cite{ethereum2014ethereum} is based on Blockchain technology, however, as of 2022, Ethereum differs from Bitcoin in that it uses Proof of Stake (PoS) as a consensus mechanism instead of PoW. In PoS, the node with the highest stake and not the highest computing power obtains the right to book-keeping, where the stake is a reflection of a node's ownership of a specific amount of currency. \cite{compareBtcEth} Therefore, it solves the problem of wasted computing power present in PoW to a certain extent and can reduce the time required to reach a consensus. However, a drawback of a stake based consensus mechanism is that it can result in centralisation in extreme cases.

Ethereum is also different from Bitcoin in that it is a programmable blockchain platform; using Smart Contracts, users on the Ethereum Platform can not only perform simple transactions, but can also create complex transactions.

The term \textit{Smart Contract} was first termed by N. Szabo who defined it as a "\textit{computerized transaction protocol that executes the terms of a contract}" \cite{szabo}. In Ethereum, a smart contract represents a deterministic, Turing-complete program which consists of a collection of code (functions) and data (state) which are deployed to the Ethereum network and run as programmed. Smart Contracts allow users on the network to define complex rules for facilitating interactions. In Ethereum, Smart Contracts are programmed using the Solidity programming language.

However, it is worth noting that Smart Contracts in Ethereum suffer from certain limitations:

\begin{enumerate}
    \item \textbf{Vulnerability Prone} \\ Smart Contract code is prone to vulnerabilities which can be costly when exploited. In a famous example, \$50 million was stolen from a DAO in an attack that exploited a concurrency-based vulnerability. A trivial error is bound to be exploited by hackers, though this limitation is true for all open source code in general.
    \item \textbf{Expensive} \\  Code in a Smart Contract running on the Ethereum network is significantly costlier per instruction than the same code running on a typical cloud server.
\end{enumerate}

Further, even though Ethereum uses PoS, it still suffers from high transaction fees similar to Bitcoin.

\subsection{IOTA}
IOTA is a cryptocurrency meant for the IoT industry. IOTA is different from the technologies listed so far in that it does not use Blockchain as its underlying ledger and instead uses the \textit{Tangle} \cite{tangle}, a directed acyclic graph (DAG) for storing transactions. Tangle is both a decentralised data storage architecture and a consensus protocol where each node in its DAG represent a transaction and the connections between nodes represent validations of the transactions. 

Unlike Bitcoin and Ethereum, IOTA removes the dichotomy between transaction miner and validator, instead requiring users who are adding transactions to the network to validate other transactions. Thus, in practice, in order to add a transaction to the network, a user needs to choose and validate two other transactions, using Hashcash as the PoW algorithm with a lowered difficulty. The rationale behind using PoW is to prevent spam transactions. Through this mechanism, IOTA is able to reduce or get rid of transaction costs and allow a much higher transaction per second count than Bitcoin and Ethereum. Further, since it relies on network users to validate transactions, as the number of users in the network increases, the validation time of submitted transactions also decreases, rendering it much more scalable than both Bitcoin and Ethereum.

However, as of the time of writing this paper, IOTA relies on a centralised coordinator to assist in adding transactions to the network and further, it does not have support for smart contracts. (Though this is planned to change in a future release). Further, since the technology is still relatively new, it is still experimental and may contain security threats. \cite{iotavuln}

\subsection{TrustChain}
A large drawback of the solutions listed so far is that none of them tackle the infrastuctural threats presented in section \ref{problems}, most notably, they are not Sybil resistant, hence, while they are useful in isolation for facilitating transactions, they cannot be used for the use case of creating trust. TrustChain \cite{trustChain} is unique in this sense, since it is a Sybil-resistant, scalable blockchain. TrustChain relies on \textit{NetFlow} a Reputation Mechanism, to calculate reputation using an interaction graph and the max-flow algorithm, thus allowing it to be Sybil resistant.

In TrustChain, each node is responsible for storing their own Blockchain and hence, transaction history. In addition to containing 
a single transaction's data, each block in the Blockchain also contains two references, one to the last block in its own Blockchain and another to the last block in the transacting party's Blockchain. While this mechanism doesn't allow the prevention of double spending, it allows detection of it since all blocks must have two outgoing and two incoming links. Further, the additional pointer to the counterparty's chain makes it hard to remove or alter blocks in one's chain, resulting in the Blockchains being tamper proof.
TrustChain blocks are also exchanged between nodes using gossiping mechanisms and hence, replicated network wide, allowing the network to be resilient against nodes going offline.

TrustChain's local storage also allows it to be highly scalable since there is no global consensus mechanism, also removing the need for transaction fees and leading to a large amount of transaction per second.

Though the lack of a global Blockchain and hence, lack of a currency earned by validators in the network could also be seen as a drawback since a large reason for growth behind popular cryptocurrencies is the investment of the users in it since they possess a vested interest in it. Further, unlike Ethereum, TrustChain does not possess a mechanism for creating complex contracts and hence, lacks flexibility in trust generation.

Table \ref{table:2} provides an overview of the DLTs covered in this section. \cite{bbdt} provides an extensive survey of how DLTs, like Blockchain are being used in Distributed Trust and Reputation Management systems.

\section{Other Mechanisms}
Besides direct reciprocity and indirect reciprocity, there are also other mechanisms that should be considered when understanding how cooperation could evolve in a decentralised network.

    \subsection{Network Reciprocity}
    While the analysis so far relies on a well-mixed population, in reality the spatial structures of social connections are not well mixed, instead certain groups interact with each other more often than others. In such a setting, it may be able to form network cluster of cooperators who help each other out resulting in a ``Network Reciprocity" which is a generalisation of ``Spatial Reciprocity". \cite{5rules}

    In their paper ``The WebEngine - A Fully Integrated, Decentralised Web Search Engine", Mario M. Kubek and Herwik Unger \cite{theWebEngine} suggest an idea idea of constructing ``content overlay networks". This involves creating social graphs with nearby and distant neighbours, where nearby neighbours are neighbours that share similar content.
    \subsection{Machine Learning based reputation systems}
    \cite{svm} suggests a reputation system that utilizes SVMs in order to establish trustworthiness of nodes in a decentralised network and demonstrates its effectiveness

\section{Future Roadmap}
Reputation systems are a powerful mechanism for preventing infrastructural threats to long term cooperation such as Sybil Attacks, however, in isolation, they possess multiple limitations that prevent them from providing a decentralised trusted ecosystem. These limitations allow Distributed Ledger Technologies to be a strong complement to reputation systems as seen in TrustChain. 

However, none of the solutions in the existing literature for generating trust attempt to tackle Social Issues inherent in decentralised systems. While in a centralised system, it is possible to solve these issues in an ad-hoc manner, decentralised systems require all rules of the system to be explicitly stated upfront and hence a ``Universal Trust Machine" would have to tackle these issues in order to create an ecosystem that can rival a centralised system.

\section{Conclusion}
This survey presents the progress of Web3 on its road towards creating a ``Universal Trust Machine", a hypothetical ecosystem where decentralised trust is generated without the aid of any trusted third parties. In order to do so, we first motivate the problem of creating cooperation in a sea of adversaries by discussing Robert Axelrod's research on the Evolution of Cooperation. Next, we clarify terminology by presenting context on the concepts of \textit{Decentralisation} and \textit{Web3}. We then present issues that such a machine would have to tackle in order to foster long term cooperation. Finally, we present contemporary technologies such as reputation systems and Distributed Ledger Technologies which in conjunction could be used to construct the desired machine. 

\bibliographystyle{IEEEtran}
\bibliography{references}

\end{document}